# Computational models for the evolution of world cuisines

Rudraksh Tuwani[1†], Nutan Sahoo[1,2†], Navjot Singh[1] and Ganesh Bagler[1*]

[1]Center for Computational Biology, Indraprastha Institute of Information Technology (IIIT-Delhi), New Delhi, India
[2]Sri Venkateswara College, University of Delhi, New Delhi, India
[*]Corresponding author: Ganesh Bagler, `bagler@iiitd.ac.in`

*Abstract*— Cooking is a unique endeavor that forms the core of our cultural identity. Culinary systems across the world have evolved over a period of time in the backdrop of complex interplay of diverse sociocultural factors including geographic, climatic and genetic influences. Data-driven investigations can offer interesting insights into the structural and organizational principles of cuisines. Herein, we use a comprehensive repertoire of 158544 recipes from 25 geo-cultural regions across the world to investigate the statistical patterns in usage of ingredients and their categories. Further, we develop computational models for the evolution of cuisines. Our analysis reveals copy-mutation as a plausible mechanism of culinary evolution. As the world copes with the challenges of diet-linked disorders, knowledge of the key determinants of culinary evolution can drive the creation of novel recipe generation algorithms aimed at dietary interventions for better nutrition and health.

## I. INTRODUCTION

Cooking is an endeavor that is unique to humans. It is ubiquitous across civilizations and has been suggested as a critical factor behind the increase in brain size of *Homo sapiens* [1]. Human affinity for cooking in the backdrop of diverse geographic, climatic, genetic, and religious influences has given rise to an array of culinary systems. Passed from one generation to the next, these systems form the core of our cultural heritage [2]. While it is generally accepted that cuisines have evolved over a period of time to optimize for human sensibilities, knowledge of the key factors that drive its evolution still evades us.

Historically, the study of dynamics of cuisines has been hindered by the interpretation of cooking as an artistic endeavor rather than a scientific one. However, recent data-driven investigations seeking divergent patterns have discovered interesting insights into the structure and organization of world cuisines. The food pairing hypothesis which theorizes that cuisines prefer combinations of similar tasting ingredients has both been refuted and confirmed [3]–[6]. Interestingly, these studies [3]–[8] have found invariant patterns in recipe size distribution and ingredient rank-frequency distribution that transcend culinary idiosyncrasies, suggesting the involvement of common evolutionary processes.

Consequently, the reproduction of these patterns has been used as the basis for comparing the plausibility of different culinary evolution hypotheses [7],[8]. Furthermore, within the purview of these comparisons, copy-mutation has emerged as the dominant theory [7], [8]. While this may indeed be true, limitations in the variety of cuisines and statistical patterns investigated as well as the lack of appropriate controls cast doubts on both the applicability and reliability of the conclusions. To address these shortcomings, we compiled a comprehensive repository of recipes from 25 distinct geo-cultural regions of the world, developed a variety of copy-mutate models along with a null model to act as the control, and compared the plausibility of these models on the basis of their ability to not only reproduce the rank-frequency distribution of individual ingredients, but also combinations of ingredients and their categories.

In the next section, we describe the data compilation procedure along with statistics pertaining to coverage and diversity of the compiled recipes. The third section explores the divergent ingredient preferences of cuisines. Then, in the fourth section, we explore the invariant patterns in the distribution of popularity of combinations of ingredients and their categories. The culinary evolution models are defined in the fifth section followed by a detailed analysis in the succeeding section.

## II. DATA COMPILATION

We compiled a total of 158544 recipes from the following recipe aggregator websites: Genius Kitchen (http://www.geniuskitchen.com) (101226), Allrecipes

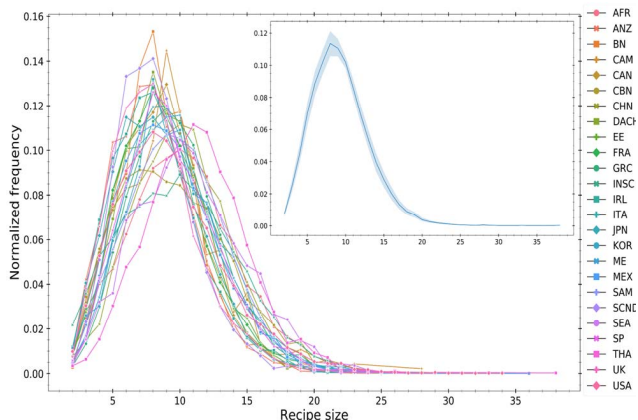

Fig. 1. Individual and aggregated (inset) recipe size distribution for the 25 world cuisines. The homogeneity of recipe size popularity is an interesting feature of culinary data.

---

[†]These authors contributed equally to this work.

TABLE I. STATISTICS OF NUMBER OF RECIPES AND INGREDIENTS AS WELL AS TOP 5 OVERREPRESENTED INGREDIENTS IN EACH WORLD CUISINE.

| Region (Code) | Recipes | Ingredients | Overrepresented Ingredients |
|---|---|---|---|
| Africa (AFR) | 5465 | 442 | Cumin, Cinnamon, Olive, Cilantro, Paprika |
| Australia & NZ (ANZ) | 6169 | 463 | Butter, Egg, Sugar, Flour, Coconut |
| Republic of Ireland (IRL) | 2702 | 378 | Potato, Butter, Cream, Flour, Baking Powder |
| Canada (CAN) | 7725 | 483 | Baking Powder, Sugar, Butter, Flour, Vanilla |
| Caribbean (CBN) | 3887 | 417 | Lime, Rum, Pineapple, Allspice, Thyme |
| China (CHN) | 7123 | 442 | Soybean Sauce, Sesame, Ginger, Corn, Chicken |
| DACH Countries (DACH) | 4641 | 430 | Flour, Egg, Butter, Sugar, Swiss Cheese |
| Eastern Europe (EE) | 3179 | 383 | Flour, Egg, Butter, Cream, Salt |
| France (FRA) | 9590 | 511 | Butter, Egg, Vanilla, Milk, Cream |
| Greece (GRC) | 5286 | 405 | Olive, Feta Cheese, Oregano, Lemon juice, Tomato |
| Indian Subcontinent (INSC) | 10531 | 462 | Cayenne, Turmeric, Cumin, Cilantro, Ginger, Garam Masala |
| Italy (ITA) | 23179 | 506 | Olive, Parmesan Cheese, Basil, Garlic, Tomato |
| Japan (JPN) | 2884 | 382 | Soybean sauce, Sesame, Ginger, Vinegar, Sake |
| Korea (KOR) | 1228 | 291 | Sesame, Soybean sauce, garlic, Sugar, Ginger |
| Mexico (MEX) | 16065 | 467 | Tortilla, Cilantro, Lime, Cumin, Tomato |
| Middle East (ME) | 4858 | 423 | Olive, Lemon juice, Parsley, Cumin, Mint |
| Scandinavia (SCND) | 3026 | 377 | Sugar, Flour, Butter, Egg, Milk |
| South America (SAM) | 7458 | 457 | Beef, Onion, Pepper, Garlic, Mushroom |
| South East Asia (SEA) | 2523 | 361 | Fish, Sugar, Soybean sauce, Garlic, Lime |
| Spain (SP) | 4154 | 413 | Olive, Paprika, Garlic, Tomato, Parsley |
| Thailand (THA) | 3795 | 378 | Fish, Lime, Cilantro, Coconut Milk, Soybean sauce |
| USA (USA) | 16026 | 592 | Butter, Sugar, Vanilla, Flour, Mustard |
| Belgium-Netherlands (BN) | 1116 | 323 | Butter, Flour, Egg, Sugar, Milk |
| Central America (CAM) | 470 | 294 | Salt, Tomato, Onion, Macaroni, Celery |
| United Kingdom (UK) | 5380 | 456 | Butter, Flour, Egg, Sugar, Milk |

(http://allrecipes.com) (16131), Food Network (https://www.foodnetwork.com) (15771), Epicurious (https://www.epicurious.com) (11022), Taste AU (https://www.taste.com.au) (7633), The Spruce (https://www.thespruce.com) (3830), TarlaDalal (http://www.tarladalal.com) (2538), My Korean Kitchen (https://mykoreankitchen.com) (198), and Kraft Recipes (http://www.kraftrecipes.com) (195). In addition to basic recipe attributes such as name, cooking procedure, and ingredient list, multi-level annotation (continent, region, and country) pertaining to the geo-cultural origin/use of the recipe was also extracted. The 'region' annotation was found to present the ideal balance between generalness and specificity and was consequently denoted as the cuisine of a recipe.

The ingredient lexicon from FlavorDB [9] was used as the base for constructing a standardized dictionary of ingredients. Specifically, 96 compound ingredients (e.g. 'tomato puree', 'ginger garlic paste' etc.) consisting of multiple individual ingredients were added to the lexicon and all the ingredients were manually assigned one of the following 21 categories: Vegetable, Dairy, Legume, Maize, Cereal, Meat, Nuts and Seeds, Plant, Fish, Seafood, Spice, Bakery, Beverage Alcoholic, Beverage, Essential Oil, Flower, Fruit, Fungus, Herb, Additive, and Dish.

Each ingredient-mention in a recipe was mapped to one of the 721 entities in our ingredient lexicon using the aliasing protocol as described in Bagler and Singh [6]. Table I presents the cuisine-wise statistics of the number of recipes and unique ingredients as well as the top 5 overrepresented ingredients (see Section III). All the cuisines are well represented in the dataset, with the average number of recipes and ingredients compiled being 6338 and 421 respectively. The largest collection of recipes is from Italy (23179) whereas the lowest is from Central America (470). These statistics highlight the broad coverage and the richness of details in our dataset. Interestingly, we found that the recipe size distribution for all the 25 world cuisines was gaussian and bounded between 2 and 38 (Fig. 1), with the average being approx. 9. Intuitively, a recipe needs to maintain a balance between complexity and simplicity to survive successive iterations of evolution. Too many required ingredients would make its propagation difficult, whereas too few would lead to it being modified easily.

III. CULINARY DIVERSITY

To probe for the differences in the ingredient preferences of world cuisines, we computed the Ingredient Overrepresentation metric. For an ingredient $i$ and region $\varsigma$, the Ingredient Overrepresentation metric $O_i^\varsigma$ was defined as:

$$O_i^\varsigma = \frac{n_i^\varsigma}{N_\varsigma} - \frac{\sum_{C=c}^{25} n_i^c}{\sum_{C=c}^{25} N_c} \quad (1)$$

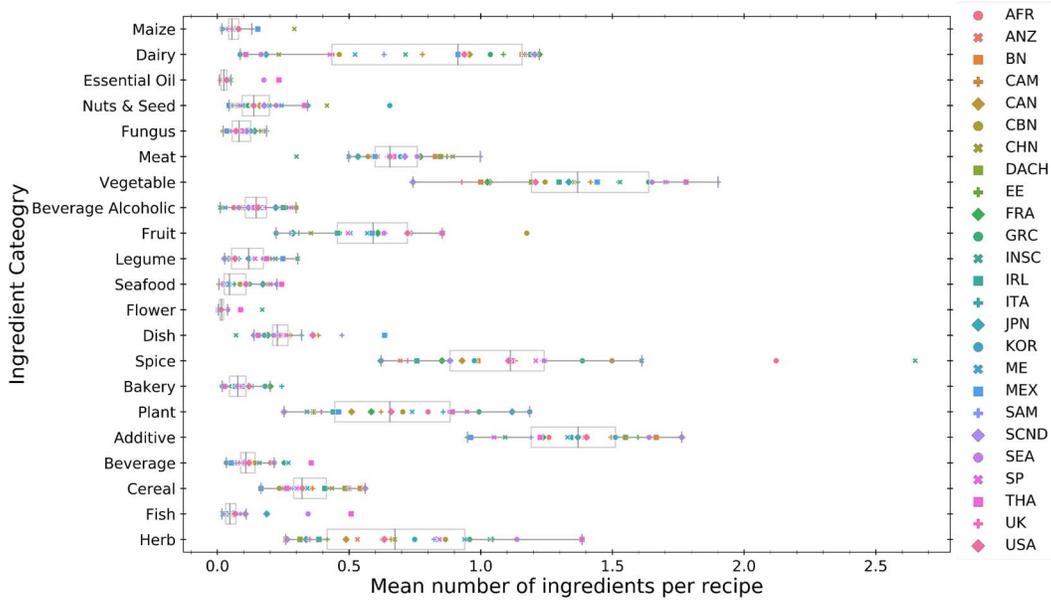

Fig. 2. Boxplots depicting the average number of ingredients used per recipe from a specific category in different world cuisines.

where $n_i^\varsigma$ is the number of recipes containing ingredient $i$ in cuisine $\varsigma$ and $N_\varsigma$ is the total number of recipes in that cuisine. $O_i^\varsigma$ is positive if the ingredient $i$ is present in a larger proportion of recipes of cuisine $\varsigma$ than across all 25 cuisines and negative otherwise. The metric quantifies the uniqueness of use of an ingredient in a specific cuisine as compared to its general use across all world cuisines. The top 5 overrepresented ingredients in each world cuisine is displayed in Table I. The diversity of world cuisines is accentuated by their unique ingredient preferences. For instance, 'fish' features prominently in South East Asian (SEA) and Thai (THA) cuisines, but it is not in the top 5 overrepresented ingredients of any other cuisine. Similarly, 'basil' is overrepresented only in Italian (ITA) cuisine.

Beyond unique ingredient preferences, the category composition of recipes also differed between distinct cuisines (Fig. 2). While all the world cuisines in-general used ingredients from Vegetable, Additive, Spice, Dairy, Herb, Plant and Fruit categories more frequently than from other categories, the average number of ingredients used from a category varied greatly. For instance, recipes corresponding to Indian Subcontinent (INSC) and African (AFR) cuisines used spices more frequently than those from Japan (JPN), Australia and New Zealand (ANZ) and Republic of Ireland (IRL). Similarly, recipes from Scandinavia (SCND), France (FRA) and Republic of Ireland (IRL) used dairy products more frequently than Japan (JPN), South East Asia (SEA), Thailand (THA), and Korea (KOR).

## IV. INVARIANT PATTERNS

The previous section demonstrated the idiosyncratic ingredient preferences of world cuisines. While the popularity of individual ingredients indeed varies from one cuisine to another, it has been shown that the pattern of ingredient popularity (rank-frequency distribution) is consistent across different regions [3]–[8]. Going beyond the level of individual ingredients, we investigated the patterns in popularity of combination of ingredients and their categories. Naturally, calculating all possible combinations would make the problem intractable. Therefore, we considered only those combinations (of size 1 and greater) which appeared in at least 5% of all recipes in a cuisine.

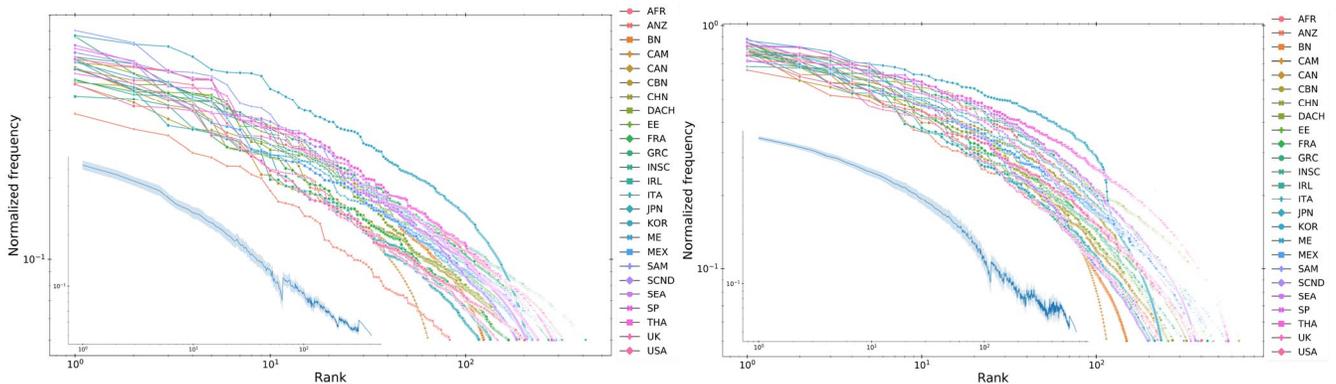

Fig. 3. Cuisine-wise and aggregate (inset) distribution of popularity of combinations of (a) ingredients and (b) ingredient categories. While the popular ingredients and ingredient categories varied between distinct cuisines, the rank-frequency (normalized by the total number of recipes) plots were homogeneous.

We found that the world cuisines had remarkably similar rank-frequency distribution of combinations of ingredient and their categories (Fig. 3). To quantify the similarity, we calculated the pairwise Mean Absolute Error (MAE) between the rank-frequency distributions of different cuisines. Specifically, the MAE between cuisines $a$ and $b$ was defined as:

$$\frac{1}{r}\sum_{i=1}^{r}(f_i^a - f_i^b)^2 \quad (2)$$

where $r$ is the lowest rank present in both the cuisines and $f_i^a$, $f_i^b$ are the normalized (by the total number of recipes in a cuisine) frequencies corresponding to the $i^{th}$ rank in cuisines $a$ and $b$ respectively. The average MAE was 0.035 and 0.052 for ingredient and category combinations respectively. In general, cuisines with small number of curated recipes (Central America, Korea etc.) had the most distinct rank-frequency distributions. If more recipes are curated for the aforementioned regions, it is possible that their rank-frequency distributions will become consistent with other cuisines.

## V. CULINARY EVOLUTION MODELS

Present-day cuisines would have evolved over time from a much smaller primitive recipe pool. Consistency in the rank-frequency distribution of combination of ingredients and their categories across different cuisines is indicative of common evolutionary processes that transcend geographical, climatic, genetic, and cultural barriers. To investigate the underlying dynamics of culinary evolution, we implemented variations of the copy-mutate model proposed by Kinouchi et al. [7]. The basic copy-mutate model with no restrictions on the choice of replacement ingredient (Copy-Mutate Random) is described in Algorithm 1. These models mimic the evolution of cuisines by incorporating duplication and alteration of recipes.

**Step 1:** Each ingredient is assigned a 'fitness' value which is randomly sampled from a $Uniform\ (0,1)$ distribution. Fitness can be interpreted as a metric quantifying the worthiness of an ingredient based on intrinsic properties such as cost, availability, and nutritional content.

**Step 2:** An ingredient pool ($I_0$) is created by randomly choosing $m$ ingredients from all the available ingredients ($I$) in a cuisine. The recipe pool $R_0$ of size $n$ is created by repeatedly sampling $\bar{s}$ ingredients (without replacement) from the ingredient pool. Here $\bar{s}$ is the average recipe size of the cuisine.

**Step 3:** At each successive iteration of the copy-mutate algorithm, we select a recipe $r$ (mother recipe) from the recipe pool $R_0$ and make a copy of it for mutation.

**Step 4:** We then randomly choose an ingredient $i$ from $r$ as well as an ingredient $j$ from the ingredient pool $I_0$ and if the fitness of $j$ is greater than that of $i$, the former replaces the latter in $r$. This process of mutating recipes is carried out $M$ times and finally $r$ is added to $R_0$.

**Step 5:** After each iteration, we calculate $\partial$ which is the ratio of the size of ingredient pool to the size of recipe pool. We also calculate $\emptyset$ by taking the ratio of the total number of ingredients to the total number of recipes of a cuisine. If $\partial \geq \emptyset$, we sample new ingredients from $I$ and add it to the ingredient pool $I_0$. The total number of recipes evolved in this manner is

---

**Algorithm 1**  Algorithm for copy-mutate model

**Input:** List of ingredients in a cuisine ($I$), average recipe size of a cuisine ($\bar{s}$), size of initial recipe pool ($n$), size of initial ingredient pool ($m$), total number of recipes in cuisine ($N$), number of mutations ($M$), and ratio of the total number of ingredients to the total number of recipes in the cuisine ($\emptyset$).

**Output:** $N$ mutated recipes

1: **for** all ingredients $i$ in $I$ **do**
2:    sample a value from $Uniform\ (0,1)$
3:    assign it to $i$
4: **end for**
5: $I_0 \leftarrow$ randomly sample (without replacement) $m$ ingredients from $I$
6: $I \leftarrow I - I_0$
7: $R_0 \leftarrow$ randomly sample $\bar{s}$ ingredients $n$ times from $I_0$
8: **for** $l = 1$ to $N - n$ **do**
9:    $\partial \leftarrow m/n$
10:    **if** $\partial \geq \emptyset$ **then**
11:      $r \leftarrow$ randomly choose a recipe from $R_0$
12:      **for** $g = 1$ to $M$ **do**
13:        sample an ingredient $i$ from $r$
14:        sample an ingredient $j$ from $I_0$
15:        **if** fitness of $j$ > fitness of $i$ **then**
16:           replace $i$ with $j$ in $r$
17:        **end if**
18:      **end for**
19:      $R_0 \leftarrow R_0 + r$
20:      $n \leftarrow n + 1$
21:    **else**
22:      choose an ingredient $p$ randomly from $I$
23:      $I_0 \leftarrow I_0 + p$
24:      $m \leftarrow m + 1$
25:      $I \leftarrow I - p$
26:    **end if**
27: **end for**

---

equal to the recipe count in the empirical data minus the size of the initial recipe pool. For normalization purposes, we create 100 such sets of random copy-mutate recipes and study the aggregated statistics.

We implemented the following derivatives of the simple copy-mutate algorithm described in the preceding paragraph which differ only in the manner of how an ingredient $j$ is chosen from the ingredient pool to replace an ingredient in the mother recipe $r$:

- Copy-Mutate Random (CM-R)
  This is the same model as the vanilla copy-mutate model described above.

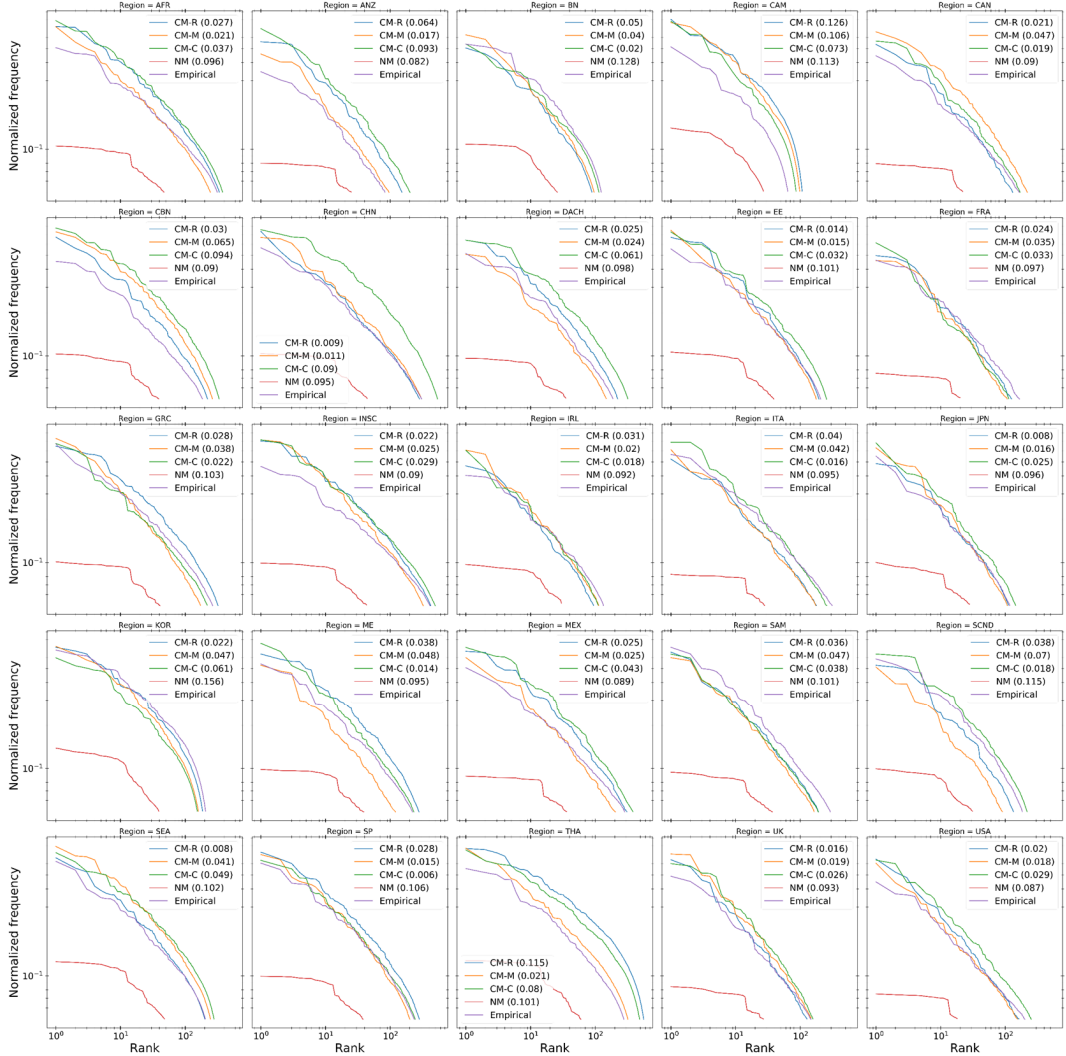

Fig. 4. Rank-frequency (normalized by total number of recipes) distribution of combinations of ingredients for all the 25 world cuisines and culinary evolution models. The MAE between the empirical and generated distribution is given in the legend.

- Copy-Mutate Category only (CM-C)
  In this model, the replacement ingredient $j$ is chosen from the same category of ingredients as $i$.
- Copy-Mutate Mixture (CM-M)
  In this model, half the time the replacement ingredient $j$ is chosen from the same category of ingredients as $i$ and otherwise it sampled from all the available ingredients.

Additionally, we implemented a Null Model (NM) wherein there are no mutations and a new recipe is created at each iteration by randomly sampling $\bar{s}$ ingredients from the ingredient pool ($I$). All the other steps remain as it is.

## VI. RESULTS

We found $m=20$, $n=I_0/\partial$, $M = 4$ (for CM-R) and 6 (for CM-C and CM-M) to consistently reproduce the empirical rank-frequency distributions of combinations of ingredients and their categories for all cuisines. In contrast, the null model was unable to replicate the empirical distributions and had high MAE across all cuisines (Fig. 4). Interestingly, the empirical rank-frequency distribution of ingredient combinations for all the copy-mutate models shows a gradual decline with rank whereas, for the null model this decline is rapid and abrupt.

The performance of copy-mutate models varied across cuisines with no discernible trends. For some regions such as Korea (KOR), Caribbean (CBN), and Japan (JPN), CM-R resulted in the lowest MAE whereas for others such as Spain (SP), Middle East (ME), Italy (ITA), and Scandinavia (SCND), CM-C had the lowest MAE. CM-M gives the best performance for Australia and New Zealand (ANZ), China (CHN), etc. Intuitively, the copy-mutate models differ in the 'creative liberty' afforded while mutating recipes. At one end of this spectrum is the CM-C model which requires the replacement ingredient to be from the same category as the ingredient to be mutated whereas at the other end is CM-R which places no such restriction. The CM-M model is in the middle, allowing cross-category mutations exactly half the time. Therefore, while the copy-mutation process may be common between cuisines, the exact mechanisms by which recipes are mutated differs with some cuisines allowing greater creative liberty than others.

We found that all the models (including null model) were able to reproduce the rank-frequency distribution of combination of ingredient categories and consequently excluded it from the analysis.

## VII. Conclusion and discussion

Culinary systems across the world have evolved over a period of time in the backdrop of complex interplay of diverse sociocultural factors including geographic, climatic, and genetic influences. Data-driven analysis can provide interesting insights into the underlying patterns that shape the structure of cuisines. In the present study, we compiled a comprehensive repository of recipes from 25 different world regions and found that despite unique ingredient preferences, the rank-frequency distributions of combinations of ingredients and their categories were consistent across all cuisines. Furthermore, our analysis suggested the role of copy-mutation process in culinary evolution based on reproduction of the aforementioned patterns.

However, our work represents only the first step in uncovering the mechanisms underlying culinary evolution. Future studies should explore the effect of variable recipe sizes, ingredient processing, and develop alternative hypotheses beyond simple copy-mutation. Furthermore, it is highly unlikely that cuisines evolved in isolation. Analogous to languages, the propagation of culinary habits would have been both vertical (time) as well as horizontal (regions).

As the world copes with the challenges of diet-linked disorders, knowledge of the key determinants of culinary evolution can drive the creation of novel recipe generation algorithms aimed at dietary interventions for better nutrition and health.

## Acknowledgment

G.B. thanks the Indraprastha Institute of Information Technology (IIIT-Delhi) for providing computational facilities and support. N. Singh, N. Sahoo are Research Interns, and R.T. is a Research Assistant in Dr. Bagler's lab (Complex Systems Laboratory) at the Center for Computational Biology, and are thankful to IIIT-Delhi for the support.

## References

[1] R. Wrangham, *Catching Fire*. Basic Books, 2010.

[2] M. Pollan, *Cooked: A Natural History of Transformation*. Allen Lane, 2013.

[3] Y.-Y. Ahn, S. E. Ahnert, J. P. Bagrow, and A.-L. Barabási, "Flavor network and the principles of food pairing.," *Sci. Rep.*, vol. 1, p. 196, 2011.

[4] A. Jain, N.K. Rakhi, and G. Bagler, "Spices form the basis of food pairing in Indian cuisine," *arXiv:1502.03815*, no. 7, p. 30, 2015.

[5] A. Jain, N. K. Rakhi, and G. Bagler, "Analysis of Food Pairing in Regional Cuisines of India," *PLoS One*, vol. 10, no. 10, 2015.

[6] G. Bagler and N. Singh, "Data-Driven Investigations of Culinary Patterns in Traditional Recipes Across the World," in *2018 IEEE 34th International Conference on Data Engineering Workshops (ICDEW)*, 2018, pp. 157–162.

[7] O. Kinouchi, R. W. Diez-Garcia, A. J. Holanda, P. Zambianchi, and A. C. Roque, "The non-equilibrium nature of culinary evolution," *New J. Phys.*, vol. 10, no. 7, p. 073020, 2008.

[8] A. Jain and G. Bagler, "Culinary evolution models for Indian cuisines," *Phys. A Stat. Mech. its Appl.*, vol. 503, pp. 170–176, 2018.

[9] N. Garg *et al.*, "FlavorDB: A database of flavor molecules," *Nucleic Acids Res.*, vol. 46, no. D1, pp. D1210–D1216, 2018.